# Strong Differential Photoion Circular Dichroism in Strong-Field Ionization of Chiral Molecules

**Authors:**
K. Fehre[*1], S. Eckart[*], M. Kunitski[*], C. Janke[*], D. Trabert[*], M. Hofmann[*], J. Rist[*], M. Weller[*], A. Hartung[*], L. Ph. H. Schmidt[*], T. Jahnke[*], H. Braun[†], T. Baumert[†], J. Stohner[‡], Ph. V. Demekhin[†], M. S. Schöffler[*2] and R. Dörner[*3]

**Affiliations:**
[*]Institut für Kernphysik Goethe-Universität Frankfurt Max-von-Laue-Str. 1, 60438 Frankfurt am Main, Germany
[†]Institut für Physik und CINSaT, Universität Kassel, Heinrich-Plett-Straße 40, 34132 Kassel, Germany
[‡]ZHAW Zurich University for Applied Sciences, Departement N, Campus Reidbach, Research Group Physical Chemistry Einsiedlerstrasse 31, 8820 Wädenswil, Switzerland

**Abstract:**
We investigate the differential ionization probability of chiral molecules in the strong field regime as a function of the helicity of the incident light. To this end, we analyze the fourfold ionization of bromochlorofluoromethane (CHBrClF) with subsequent fragmentation into four charged fragments and different dissociation channels of the singly ionized methyloxirane. We observe a variation of the differential ionization probability in a range of several percent. Accordingly, we conclude that the helicity of light is a quantity that should be considered for the theoretical description of the strong field ionization rate of chiral molecules.

Circular dichroism (CD) effects depict the difference in the absorption strength between left and right handed circularly polarized light occurring for the two enantiomers of a chiral substance. CD effects have received great attention for many years with a variety of possible applications in, for example, the fields of absorption spectroscopy [1] or fluorescence spectroscopy [2]. Since the probability of absorption is directly linked to the probability of photoionization [3], an occurring CD can lead to observable differences in the absolute ion yield [4]. CD effects are usually very small; their theoretical description relies on light-matter interaction beyond the electrical dipole approximation (see Supplemental Materials), which is a general feature for all scalar observables connected to the scope of CD [5]. Thus, one promising route to enhance the visibility of CD effects is to examine vectorial observables such as electron emission angles (electron momenta) rather than rates or yields. One prime example showing the strength of such an approach is the photoelectron circular dichroism (PECD), which is the normalized change of photoelectron angular emission distributions from chiral molecules upon inversion of the light helicity. The PECD displays signal strengths increased by orders of magnitude [6,7–9] as compared to the emission angle-integrated CD. In the present paper, we show that ion momenta are a powerful observable increasing chiral sensitivity by an order of magnitude compared to the scalar photoion count rate. The differential ionization probability of molecules can strongly depend on the orientation of the molecule with respect to the laser, which has been shown experimentally [10] (see Fig. 2) and can now be routinely modeled by standard quantum chemistry program packages [11]. The chiral effect discussed in this paper, however, is the influence of the helicity of the light on the ionization probability that is resolved on the fragmentation direction of the molecule with respect to the light propagation direction.

While typical enhancements of the photoion circular dichroism PICD for a fixed molecular orientation are rather small for single photon ionization [12], a recent experiment fosters the hope for much bigger effects in the strong field case [13]. In this experiment, an achiral molecule was excited and its chiral fragmentation detected. While the light's helicity had only a very slight influence on the total ion yield of an enantiomer, the differential PICD led to a signal amplification of up to two orders of magnitude. However, the observed changes in the differential ionization probability were inextricably linked to the selective generation of one of the two enantiomers. Hence, the observation of a strong differential PICD employing a *real* chiral molecule is still pending.

The experiments reported here have been performed using the ion arm of a COLTRIMS (Cold Target Recoil Ion Momentum Spectroscopy) [14] reaction microscope as described in previous publications [9,13,15]. Two independent experiments were performed using bromochlorofluoromethane (CHBrClF) and methyloxirane ($C_3H_6O$). The ionization of the CHBrClF / methyloxirane molecules was induced by focussing short, intense laser pulses (f = 60 mm, 40 fs, beam diameter 8.2 / 4 mm (FWHM), central wave length 800 nm, 2.5 W / 0.3 W), generated by a Ti:Sapphire regenerative amplifier (KMLabs Wyvern 500), resulting in a focal intensity of $2.5 \cdot 10^{15}$ W/cm² / $6.9 \cdot 10^{13}$ W/cm² onto supersonic gas jet jets of either a racemic mixture of CHBrClF or enantiopure $C_3H_6O$. High-efficiency microchannel plates were used [16]. With an ionization potential of 10.25 eV [17] (methyloxirane), the above given parameters result in a Keldysh-parameter of γ=0.495. During the data acquisition, a motorized stage switched the helicity of the light every 3 minutes to ensure identical experimental conditions for the measurements with left- and right-handed polarized light (LCP and RCP). Generation of LCP and RCP light was done by using a set of a half- and a quarter-wave plates. The angles of the plates were optimized by minimizing the ionization yield of argon. The final ionization yields of argon for LCP and RCP light were the same within an accuracy of 1 %. The gas jet was produced by expanding CHBrClF / methyloxirane vapour (vapor pressure at room temperature is approx. 600 / 588 mbar, respectively) through a nozzle of 30 μm diameter into vacuum and differentially pumped / shaped by three stages with (adjustable) apertures. Chlorine and bromine mainly occur as two isotopes: $^{35}$Cl (75.77 %) and $^{37}$Cl (24.23 %), $^{79}$Br (50.69 %) und $^{81}$Br (49.31 %). Hence, there are four possible combinations for the masses of the four fragments in the observed reaction. Assuming that the isotope has no influence on the PICD (isotopes are isoelectric, the greatest observable influence results from slight variation in the bond length due to the different masses), the four combinations for the isotopes were analyzed individually and combined at the end of the analysis in order to increase the statistics of the dataset.

In the present experiment we examine the four-body-fragmentation of CHBrClF → $CH^+ + Br^+ + Cl^+ + F^+ + 4e^-$. We ionize molecules from a racemic sample and determine the handedness of each individual molecule from the triple-product of three of the four momentum vectors of the fragments measured in coincidence[18]. E.g., $\vec{p_F} \cdot (\vec{p_{Cl}} \times \vec{p_{Br}})/(|\vec{p_F}| \cdot |\vec{p_{Cl}} \times \vec{p_{Br}}|) < 0$ (> 0) indicates the ionization of the S (R) enantiomer. $\vec{p_F}, \vec{p_{Cl}}$ and

$\vec{p_{Br}}$ are the measured momentum vectors of the singly charged fluorine-, chlorine- and bromine-ions.

A chemical synthesis of chiral molecules without specific asymmetric synthesis aspects of enantio-separation results in a perfect racemate; the ratio between R and S enantiomers is 1:1. Therefore, both enantiomers are subject to identical experimental conditions and provide a perfect crosscheck for the measured circular dichroism in the ion yield. The measurements with the two light helicities show a normalized difference in the $\text{PICD} = \frac{R_{RCP} - R_{LCP}}{R_{RCP} + R_{LCP}} = (0.094 \pm 0.046)\,\%$ in the fourfold fragmentation of CHBrClF for both enantiomers (integrated over all molecular orientations). $R_{LCP}$ ($R_{RCP}$) indicate the measured count rate of the measured four-body fragmentation channel of the R enantiomer for LCP (RCP). In order to compensate for slight differences in the measuring time, the totals of the count rates for $R_{RCP}$ and $S_{RCP}$ ($R_{LCP}$ and $S_{LCP}$) were normalized to one. Performing the data analysis on the mirror image molecule yielded the exact same result (($-0.094 \pm 0.046)\,\%$) with the expected change of sign.

Our fourfold coincidence measurement of ion momenta allows us to inspect the data on different levels of detail. The most global and least differential level is the aforementioned ion count rate depicting a PICD of only $0.094\,\%$. A more differential and yet very basic vectorial observable is the PICD as a function of the direction of one of the ion momenta with respect to the light propagation axis. This is the ion equivalent to angular resolved PECD.

We choose the Br$^+$ ion for that purpose and plot in Fig. 1 the PICD, defined as $\text{PICD}(\vec{p_{Br}}) = \frac{R_{LCP}(\vec{p_{Br}}) - S_{LCP}(\vec{p_{Br}})}{R_{LCP}(\vec{p_{Br}}) + S_{LCP}(\vec{p_{Br}})}$, as a function of the cosine of the angle between the Br$^+$ momentum and the light propagation axis (x-direction). Please note that we use the normalized difference between the enantiomers to calculate the PICD, as this representation shows the forward / backward asymmetry, as it is known from experiments on the photoelectron circular dichroism [7,8,19]. However, for (partially) oriented molecules, the differential PICD shows a different pattern comparing the two light helicities or the two enantiomers. A detailed discussion is attached in the Supplemental Material. We find a PICD of up to 1 % for a Br$^+$ ion emission along the light propagation inverting its sign for the opposite Br$^+$ emission direction or light helicity.

As the next step we inspect the fully differential PICD incorporating all ionic fragments. For this purpose, we switch to a molecular coordinate frame. In this system, the momentum of the CH$^+$ fragment defines the z'-axis, the Br$^+$ fragment emission direction defines the z'y'-plane and points in the positive y'-direction. All ionic fragment emission directions as well as the light propagation direction are transformed into the molecular system. The result, converted into the angles of the spherical coordinates, is shown in Fig. 2.

Each point in the map in Fig. 2(a) corresponds to a distinct light propagation direction in the molecular frame; the light's polarization plane is perpendicular to this direction. For example, as expected, the ionization probability is high whenever the halogen atoms lie within the polarization plane.

Now we can turn to the investigation of the differential PICD. Fig. 3 shows the corresponding normalized difference between the ionization with LCP and RCP of CHBrClF and the aforementioned 4-fold fragmentation channel. PICD values of up to 10 % are observable and thus show signal strength that is two orders of magnitude larger than the CD in the ion yield.

The PICD patterns show in first approximation a change of sign in the PICD for molecular orientations in which the light in the molecular system impinges on the molecule from opposite directions. In the presented PICD maps, molecular orientations corresponding to the anti-symmetry are linked by a change of sign in $\cos\theta$ and a translation of $\pm 180°$ in φ. For small values of $\cos\theta$ there is a deviation of this anti-symmetry in the PICD pattern.

As the definition of a joint molecular system is not possible for the two enantiomers of a chiral molecule, is initially not clear which points on the two PICD maps are to be compared. At first glance, the two PICD patterns of the enantiomers connected by a mirror operation (mirror plane located between the panels). A closer look suggests a different symmetry relationship. The negative contribution in Fig. 3(a) at φ = -100° is at $\cos\theta$ = +0.2, and in Fig. 3(b) it is at about φ= 100° and $\cos\theta$ = -0.2. Therefore, a point inversion at φ = 0 and $\cos\theta$ = 0 seems to be a more suitable symmetry operation to connect the PICD patterns of the two enantiomers.

Fourfold ionization as in the case of CHBrClF is a rare event. In most interactions with a laser pulse, only one electron is emitted. Thus, for analytical purposes it is desirable to study PICD also in single ionization events. In many such cases the molecule still fragments fast in a neutral and a charged fragment, which makes PICD as function of the fragment emission direction accessible [20]. Motivated by the significant increase found for CHBrClF can hope to utilize this observable to enhance the PICD strength compared to the weak signal from merely detecting the ion count rate.

Please note that the COLTRIMS-technique used in the presented experiment separates the time of flight direction and the forward / backward direction with respect to the light propagation direction. We inspect the PICD in a 2-dimensional representation showing the position of impact of a particle on the detector versus the time-of-flight. Such a histogram is shown in Fig. 4 employing methyloxirane as a target. However, the differential PICD, in which only one molecular axis is resolved, is already accessible in a time of flight measurement, if the laser propagates not perpendicularly to the extraction direction of the time-of-flight spectrometer.

Figure 4 depicts, that the fragments under consideration may overlap in their flight times due to the momentum they received in the dissociation process. Thus, their masses cannot be unambiguously assigned. We therefore investigate the PICD as a function of the ion's time of flight (TOF). The dashed lines in Fig. 4 indicate the mean TOF of the fragments with $\frac{m}{q} = 43, 42, 41, 40, 39$ and $38$ amu/au. While $\frac{m}{q} = 43$ indicates the singly charged ring fragment $C_2H_3O$, the other mass to charge ratios can be reached via several fragmentation pathways. Different data recording times or imperfect circular polarizations directly influence the CD. Differences in the laser intensity can even change the weighting between the individual fragmentation channels [21]. It turns out, that these sources of error largely compensate in the displayed difference of normalized differences [19]. A small constant offset was subtracted to compensate long-term laser drifts between the measurements of the two enantiomers (details in the Supplemental Material). This demonstrates that for the CD in the ion yield a sensitive reference is needed. The CD signals presented in Fig. 4(a) are very small and in this case are not (much) larger than the experimental uncertainty. The statistical error is smaller than the estimated systematic errors. Therefore, instead of error bars we present a corresponding discussion in the Supplemental Material.

As before, the differential quantities show far stronger signals (Figs. 4(b) and (c)). In Fig. 4(b), the same value as in (a) is displayed with the additional condition that the ionic fragment impinged the detector in or against the direction of light propagation [22].

The PICD as a function of the TOF and the position of impact on the ion detector is shown in Fig. 4(c). Before calculating the normalized difference, all four contributions from the two light helicities and enantiomers were normalized to 1 for each slice in the time-of-flight. As a result, the integral CD from the ion yield disappears and Fig. 4(c) only contains the influence of the CD in the differential ionization probability. Fig. 4(c) shows the PICD in this TOF region to be anti-symmetric in forward and backward direction with a signal that is more than five times stronger than the CD signal (see Fig. 4(a)). Interestingly, the strong dissociation channel for $\frac{m}{q} = 43$ does not show a significant differential PICD. Presumably, this is due to a delay between the instant of ionization and the dissociation process itself. If the dissociation time is comparable or longer than the rotational period of the molecule, the direction of the ion momentum vector has no correlation with the molecular orientation at the instant of ionization. As confirmation of this hypothesis, no correlation between the measured ion momentum vector and the electron emission direction was found.

Thus, the PICD in the single ionization of methyloxirane confirms an amplification of the signal, if a vectorial observable is examined instead of the integral ion yield. This increase is comparable to the presented case of the four-fold ionization of CHBrClF. This trend suggests that here the dependence on the helicity of the light can influence the double differential ionization probability in a few percent range.

In conclusion, we have shown that in the strong field regime for a chiral molecule that is oriented in space, the single and multiple ionization probability significantly depends on the helicity of the light. We have found changes in the ionization probability of up to 10 %. This is almost two orders of magnitude larger than what is typically found for randomly oriented molecules. Our data show that the sign of the observed effect almost completely inverts upon mirroring of the geometry, which is the reason, why the remaining effect, which survives averaging over all orientation of the fragments is so small. From a technical perspective, our findings suggest that the enantioselectivity of ion detection can be much enhanced if in addition to the mass of the ion also one momentum component along the light propagation direction is detected. As we have shown this is easily feasible by adding a position sensitive detector to a time-of-flight mass spectrometer. Thus, we find that the influence of the light's helicity in strong field ionization of chiral molecules differs drastically from the theoretical predictions for the case of single ionization by one photon. In addition, the differential PICD with the reported signal strength contributes significantly to the differential ionization probability in the strong field regime, which should be considered in future theoretical models.

**Acknowledgments:**
This work was funded by the Deutsche Forschungsgemeinschaft (DFG) — Project Number 328961117 — SFB 1319 ELCH (Extremelight for sensing and driving molecular chirality). K.F. and A.H. acknowledge support by the German National Merit Foundation. M. S. thanks the Adolf-Messer foundation and J. S. thanks ZHAW for financial support. Discussions with L. B. Madsen are gratefully acknowledged.

**Author Information**

Correspondence and requests for materials should be addressed to [1]K. F. (fehre@atom.uni-frankfurt.de), [2]M. S. S. (schoeffler@atom.uni-frankfurt.de) or [3]R. D. (doerner@atom.uni-frankfurt.de). The authors declare no competing interests.

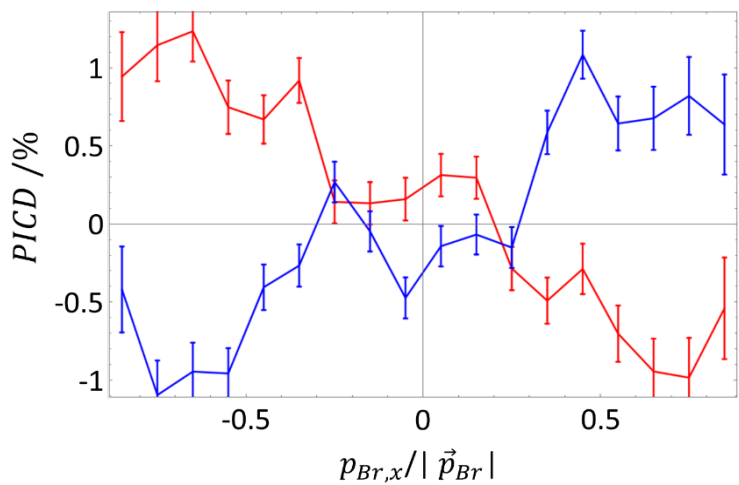

**Fig. 1 Differential PICD as normalized difference between the enantiomers for CHBrClF as function of the angle between the bromine ion momentum and the direction of light propagation.** The red curve reflects the measurement results for LCP, the blue for RCP. The error bars indicate the statistical error. Please note, for (partially) oriented molecules, the differential PICD shows a different pattern comparing the two light helicities or the two enantiomers. A detailed discussion is attached in the Supplemental Material.

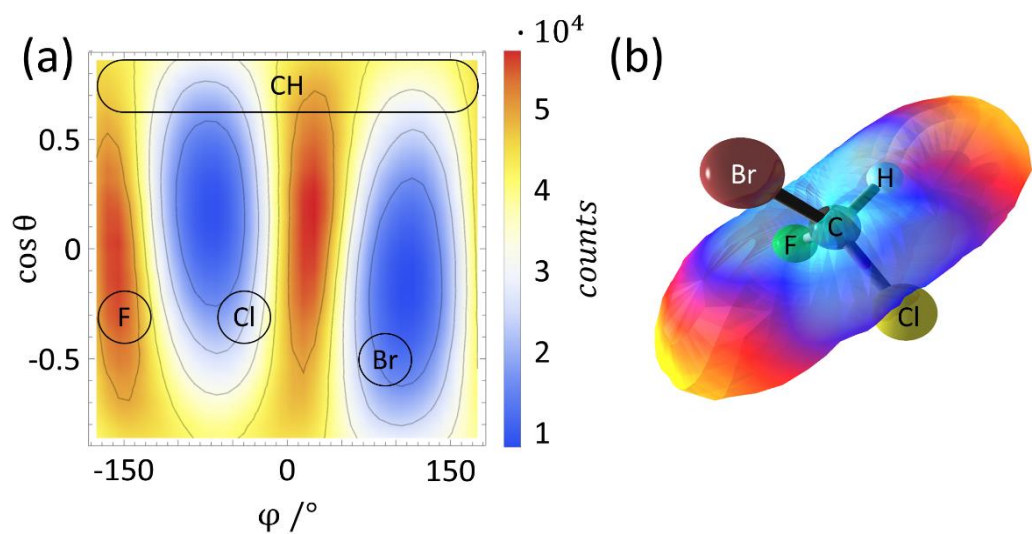

**Fig. 2 Differential ionization probability for R-CHBrClF and LCP.** (a) Each point in the graph shows the number of measured events for a direction of light propagation in spherical coordinates in the molecular system. The position of the molecule in the selected molecular system is indicated by the position of the fragments' momenta. (b) The colored sphere represents the differential count rate in the molecular system. The count rate is represented by the distance from the C-atom and the color.

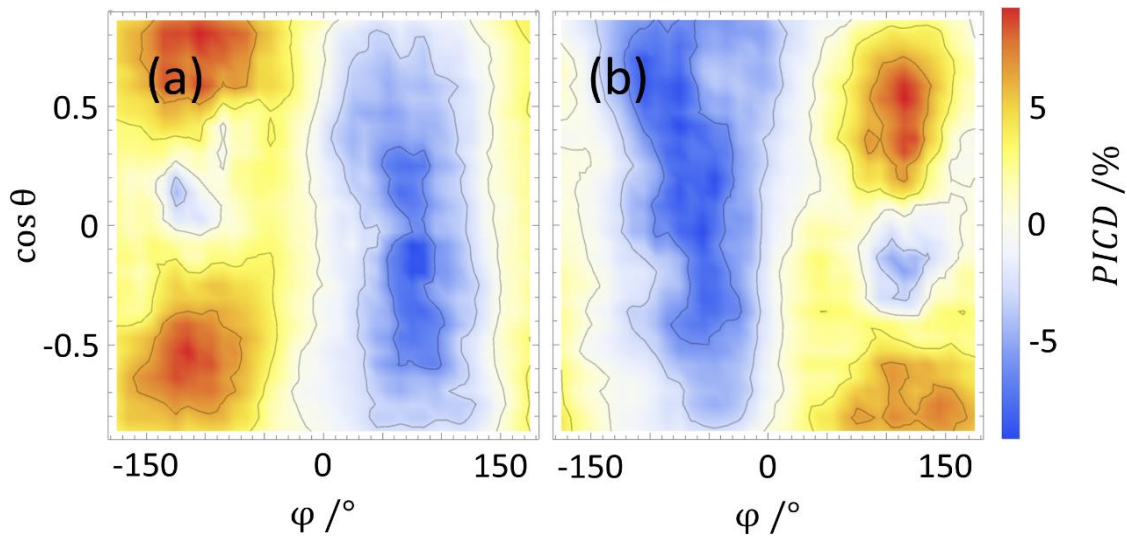

**Fig. 3 Differential PICD in the four-body fragmentation of CHBrClF.** (a) $PICD(\varphi, cos(\theta)) = 100 * \frac{S_{RCP}(\varphi,cos(\theta)) - S_{LCP}(\varphi,cos(\theta))}{S_{RCP}(\varphi,cos(\theta)) + S_{LCP}(\varphi,cos(\theta))}$ for the S enantiomer. (b) PICD as in (a) for the R enantiomer. For small values of $cos\theta$ there is a deviation from the antisymmetry in the PICD pattern.

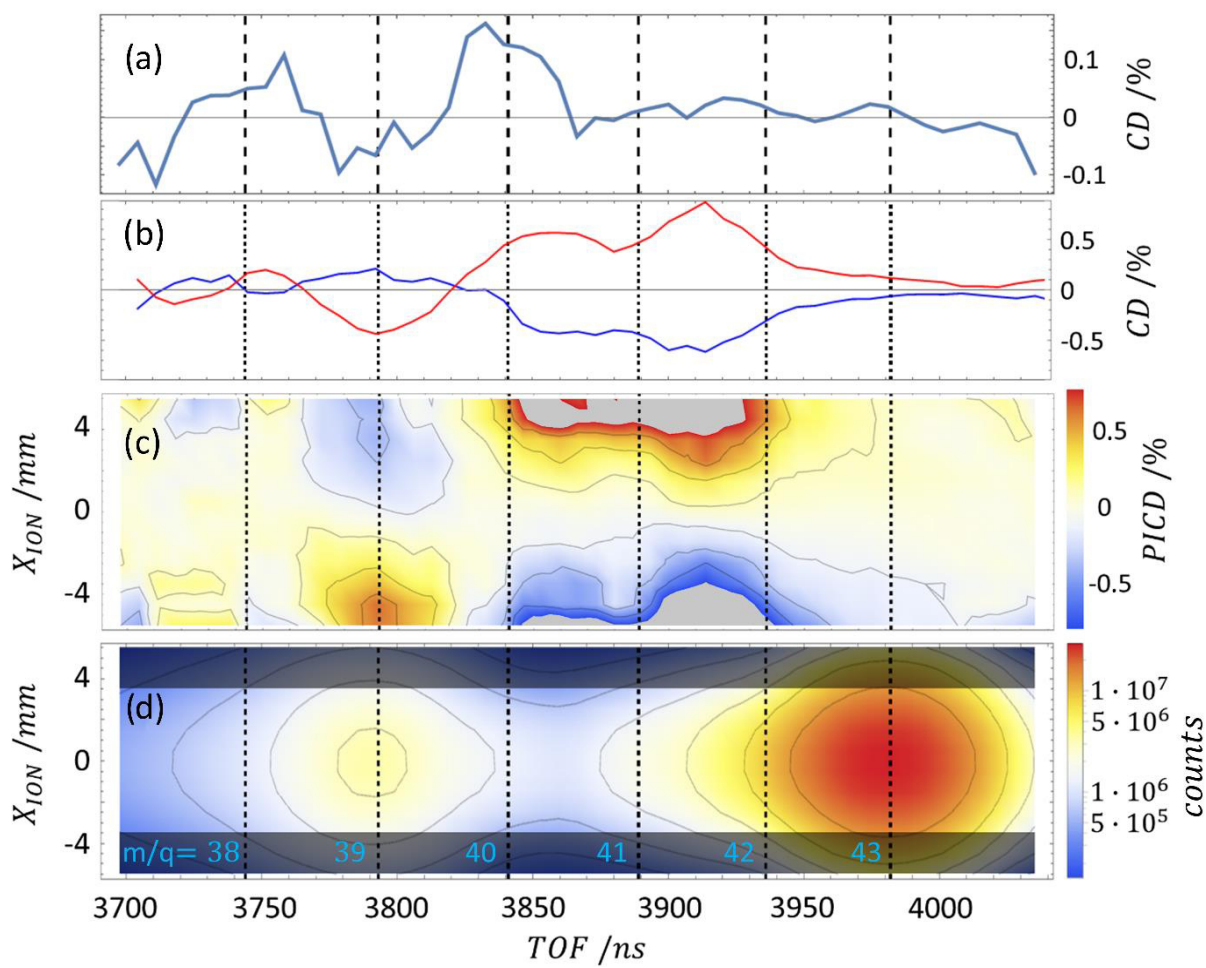

**Fig. 4 Circular dichroism in the single ionization of methyloxirane.** (a) Circular dichroism in the ion yield calculated by $CD(TOF) = 50 \cdot \frac{R_{LCP}(TOF) - S_{LCP}(TOF)}{R_{LCP}(TOF) + S_{LCP}(TOF)} - 50 \cdot \frac{R_{RCP}(TOF) - S_{RCP}(TOF)}{R_{RCP}(TOF) + S_{RCP}(TOF)} - 0.4$. Different measuring times and the smallest errors directly influence the CD. The applied small correction cannot be determined from the available data. A detailed discussion of the errors can be found in the SI. (b) As in (a) with the additional condition, that the location of the ion hit on the detector in the light propagation direction is larger (smaller) than 3.5 mm (-3.5 mm), represented by the red (blue) line. With this condition, a subset of molecular orientations is selected. (c) $PICD(X_{ION}, TOF) = 50 \cdot \frac{R_{LCP}(X_{ION}, TOF) - S_{LCP}(X_{ION}, TOF)}{R_{LCP}(X_{ION}, TOF) + S_{LCP}(X_{ION}, TOF)} - 50 \cdot \frac{R_{RCP}(X_{ION}, TOF) - S_{RCP}(X_{ION}, TOF)}{R_{RCP}(X_{ION}, TOF) + S_{RCP}(X_{ION}, TOF)}$ as a function of the ionic TOF and $X_{ION}$ /mm the position of impact onto the detector of the ion in the direction of light propagation. The count rate for each TOF and for each enantiomer and each helicity is normalized to one, whereby only the differential signal is represented; different integral ionization probabilities play no role in this representation. (d) Absolute ion count rate as a function of the ion TOF and position of impact on the ion detector. The two helicities of the light and the two enantiomers were added together in this representation. The transparent areas indicate the selection in (b). The vertical lines and their labels show the average TOF of the ions for the stated mass-to-charge ratios.

# Supplemental Material

# Strong Differential Photoion Circular Dichroism in Strong-Field Ionization of Chiral Molecules

**Authors:**

K. Fehre[*,1], S. Eckart[*], M. Kunitski[*], C. Janke[*], D. Trabert[*], M. Hofmann[*], J. Rist[*], M. Weller[*], A. Hartung[*], L. Ph. H. Schmidt[*], H. Braun[†], T. Baumert[†], J. Stohner[‡], T. Jahnke[*], Ph. V. Demekhin[†], M. S. Schöffler[*] and R. Dörner[*,2]

**Affiliations:**

[*]Institut für Kernphysik Goethe-Universität Frankfurt Max-von-Laue-Str. 1, 60438 Frankfurt am Main, Germany

[†]Institut für Physik und CINSaT, Universität Kassel, Heinrich-Plett-Straße 40, 34132 Kassel, Germany

[‡]ZHAW Research Group Physical Chemistry Einsiedlerstrasse 31, 8820 Wädenswil, Switzerland

**Discussion of systematic errors in Fig. 4**

Without resolving the molecular orientation, the same PICD signal is expected when comparing the two light helicities or enantiomers. However, as was already emphasized in the supplementary material of a previous publication [1], the respective comparisons are subject to different experimental errors. The helicity of the light was interchanged every 3 min with the help of a motorized stage, whereby both light helicities are subject to the same long-term laser drifts. Even with great care, the LCP and RCP are not perfectly mirror-symmetric, which has a direct effect on differences in the ion count rate. In addition, identical focus geometries cannot be guaranteed. The different breakup channels also show an individual dependence of their ionization probability on the laser intensity. Therefore, an achiral reference for the count rates cannot be used to make a statement about the absolute PICD zero.

Although the two enantiomers have perfect mirror symmetry, they cannot be measured quickly one after the other with the described experimental setup, since otherwise the samples would be contaminated. The long-term laser drifts (intensity and pointing) thus influence the observed ionization rate.

Fig. S1(a) shows the PICD calculated for R and S enantiomer calculated by the normalized difference of the light helicities. A striking feature that is common to the course of both the R enantiomer (green and red line) and the S enantiomer (blue line) is a similar, strong dependence on the ionic TOF. For a better comparison, the course for the R enantiomer (green) was reduced by - 0.8 (red). This strong TOF-dependence, which is the same for both enantiomers, can be attributed to the imperfect mirror symmetry between LCP and RCP, the offset between R and S enantiomer presumably to a small laser drift between the measurement with the S and the R enantiomer. The average PICD shown in Fig. 4(a) is calculated from the normalized difference in light helicities for R and S enantiomer [2]. In this representation, the discussed experimental errors cancel out to a certain extent. The measurement of the integral PICD is therefore not only a challenge due to the small signals, but also because a suitable reference for the PICD zero must be found for each fragmentation channel.

The differential analysis in Fig. 4(c) shows a solution for both hurdles when measuring the PICD. First, the signal strength of the differential PICD grows stronger than the statistical error. The latter only grows because the events are no longer viewed only as a function of the TOF, but also as a function of the point of impact onto the detector. On the other hand, there is no need for a reference for the absolute count rate. Since we are only interested in the relative ionization probability in the forward or backward direction in this diagram, each cut in the TOF can be normalized individually to one. Fig. S1(b) and (c) show the antisymmetry of the differential PICD in the forward or backward direction for the R and S enantiomer. The exception here is the S enantiomer at about $TOF = 3875\ ns$ and $X_{ION} = -4\ mm$. It would be conceivable that the higher statistics at $TOF = 3800\ ns$ for larger momenta from the dissociation process and negative PICD overlap with the PICD's forward and backward asymmetry.

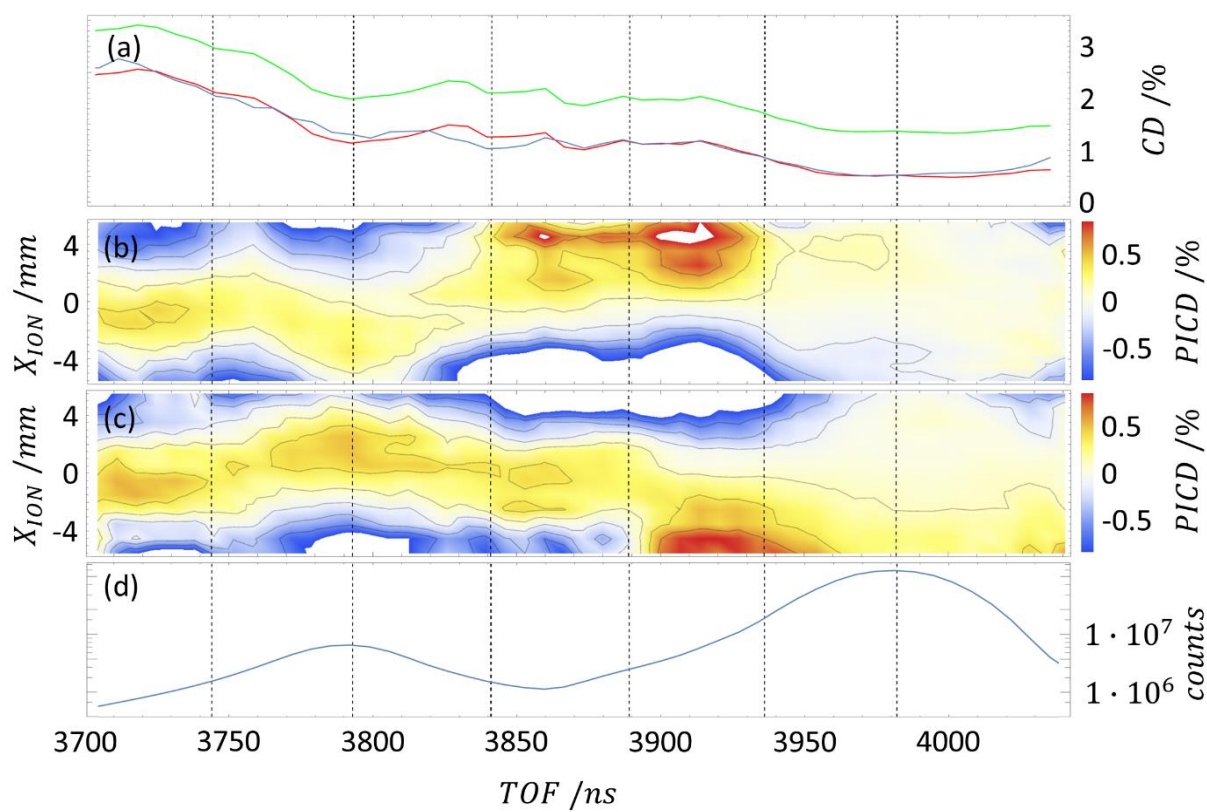

**Fig. S1 Circular dichroism in the single ionization of methyloxirane. (**a) Circular dichroism in the ion yield calculated by $CD(TOF) = 100 * \frac{R_{RCP}(TOF) - R_{LCP}(TOF)}{R_{RCP}(TOF) + R_{LCP}(,TOF)}$ for the R enantiomer (green line, and – 0.8 for the red line for better comparison). The S enantiomer is calculated by $CD(TOF) = 100 * \frac{S_{RCP}(TOF) - S_{LCP}(TOF)}{S_{RCP}(TOF) + S_{LCP}(,TOF)}$ (blue line). (b) $PICD(X_{ION}, TOF) = 100 * \frac{R_{RCP}(X_{ION},TOF) - R_{LCP}(X_{ION},TOF)}{R_{RCP}(X_{ION},TOF) + R_{LCP}(X_{ION},TOF)}$ as a function of the ionic TOF and $X_{ION}\ /mm$ the position of impact onto the detector of the ion in the direction of light propagation for the R enantiomer. The count rate for each TOF and each helicity is normalized to one. (c) Same as B for the S enantiomer. (d) Counts as function of the TOF. The vertical lines show the average TOF of the ions for the in Fig. 4 stated mass-to-charge ratios.

**Differences in the PICD when comparing the enantiomers or light helices**

If the molecule is (partially) fixed in space, it usually makes a difference whether the normalized difference is formed between the two light helicities or the enantiomers. A comparable observation has already been made for the photoelectron circular dichroism when examining it with elliptically polarized light[3]. This very different PICD as function of the molecular orientation can be seen by comparing Fig. S2 with Fig. 1. In Fig. 1, the normalized difference is calculated from the enantiomers, in Fig. S2 from the light helicities.

By definition, no common coordinate system exists for the two enantiomers of a chiral molecule: For the mirror image enantiomer, the $F^+$ and the $Cl^+$ fragment emission directions swap their location in the molecular frame. If one enantiomer investigated with the same definition of the coordinate system as the other, the location of the maxima of the ionization probability differs and a normalized difference between the two graphs leads to differences of over 60 %.

From Fig. 3 we can see the symmetry that connects the PICD-Maps for the two enantiomers: The mirror image enantiomer does not lead to a change of sign in every point in the PICD map. The points in the map for the one enantiomer can be connected by a point inversion at $\varphi = 0$ and $\cos\theta = 0$ to those of the other enantiomer. The shape of Fig. S2 can be derived from Fig. 2 and Fig. 3: For every bin in Fig. S2 the region with the corresponding intermediate angle from the location on the map and position of the $Br^+$- ion has to be integrated and weighted with the differential ionization probability from Fig. 2. Accordingly, the symmetry property for the PICD signal that can be expected is that a reversal in the direction of the ion is equivalent with the exchange of the enantiomer. This symmetry can be seen in Fig S2.

The situation is different in Fig. 1, where the normalized difference is calculated from the enantiomers, whereby a mixed coordinate system is considered: That of the R and that of the S enantiomer. A similar mixing occurs when the difference between the two PICD curves in Fig. S2 is formed; the result almost corresponds to that presented in Fig. 1. This transfer explains the forward-backward antisymmetry that can be observed in Fig. 1.

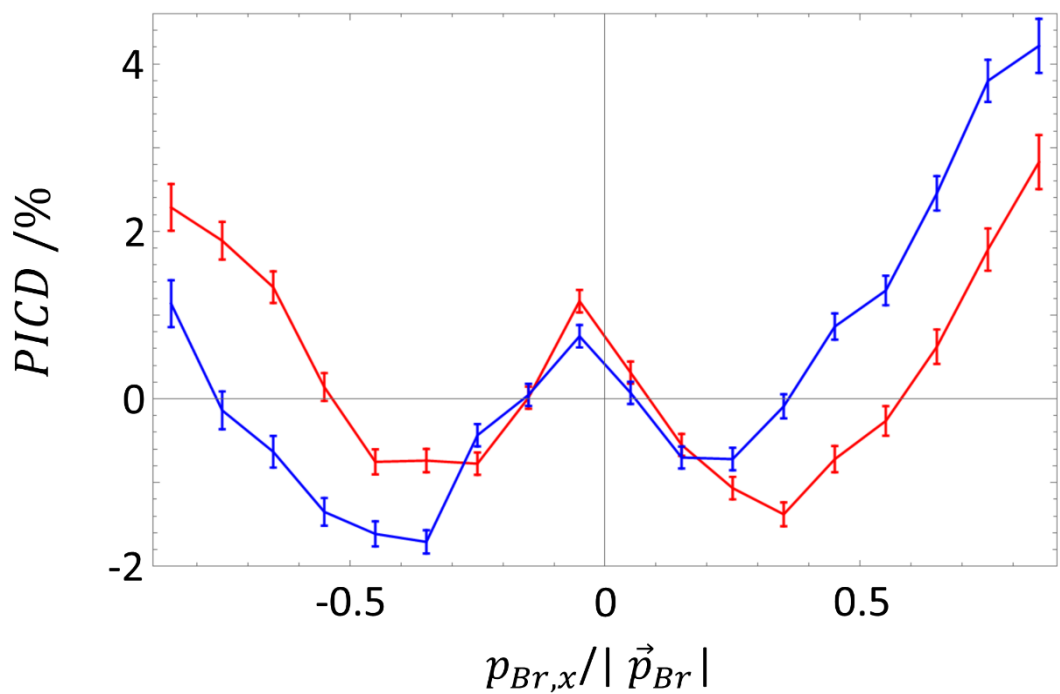

**Fig. S2 Differential PICD as normalized difference between the light helicities as function of the angle between the bromine ion momentum and the direction of light propagation.** The red curve reflects the measurement results for the R enantiomer, the blue for the S enantiomer. The error bars indicate the statistical error.

**Explanation on the PICD in single photon ionization**

For ionization by a single photon Cherepkov has shown that this possible enhancement of photoion circular dichroism (PICD) for fixed molecular orientation is rather small[4]. This is because PICD in the one-photon ionization of oriented chiral molecules relies on the interferences between the electric-dipole with magnetic-dipole as well as electric-dipole with electric-quadrupole interactions, and only the former interference survives averaging over molecular orientations and contributes to the CD[4].